 \documentclass[final,1p,times,twocolumn]{elsarticle}

\usepackage{amssymb}
\usepackage{url}

\newcommand\op[1]{\mathop{\rm #1}\nolimits}

\newcommand{\weg}[1]{}

\newtheorem{theoremc}{Theorem}
\newtheorem{rk}[theoremc]{Remark}

\usepackage{fullpage}
\usepackage{color}
\journal{Physica A}

\begin{document}

\begin{frontmatter}


%

 \title{Assessing market uncertainty by means of a time-varying intermittency parameter for asset price fluctuations}


\author[label1]{Martin Rypdal\corref{cor1}}
 \ead{martin.rypdal@uit.no}
 \address[label1]{Department of Mathematics and Statistics, University of Troms{\o}, N-9037 Troms{\o}, Norway.}
 \author[label2]{Espen Sirnes}
\address[label2]{Troms{\o} University Business School, University of Troms{\o}, N-9037 Troms{\o}, Norway}
\author[label1]{Ola L{\o}vsletten}
 \cortext[cor1]{Corresponding author}
 \author[label3]{Kristoffer Rypdal}
\address[label3]{Department of Physics and Technology, University of Troms{\o}, N-9037 Troms{\o}, Norway}

\begin{abstract}
 Maximum likelihood estimation applied  to high-frequency data allows us to quantify intermittency in the fluctuations of asset prices.  From time records as short as one month these methods permit  extraction of  a meaningful intermittency parameter $\lambda$ characterising the degree of volatility clustering of asset prices. We can therefore study the time evolution of volatility clustering and test the statistical significance of this variability. By analysing data from the Oslo Stock Exchange, and comparing the results with the investment grade spread, we find that the estimates of $\lambda$ are lower at times of high market uncertainty.   
\end{abstract}

\end{frontmatter}


\section{Introduction} \label{introduction}
In the 1990s it was discovered that multifractal stochastic processes are useful as models for financial time series  \cite{Ghashghaie:1996wa,Mandelbrot:1997wz}\color{black}. The advantage of these processes is that they naturally combine uncorrelated returns with long-range volatility dependence, and therefore provide accurate models without over-parametrization. Several comparison tests have shown that multifractals out-perform generalized autoregressive conditional heteroskedasticity (GARCH) type models as descriptions of asset prices, currency exchange rates and short-term interest rates \cite{Calvet:2004ty,Rypdal:2011vv}\color{black}.  

A stochastic process $X(t)$ is denoted multifractal if it exhibts stationary increments and its structure functions $S_q(t) = \mathbb{E}[|X(t)|^q]$ are power-laws as functions of $t$, i.e., $S_q(t) \sim t^{\zeta(q)}$. The scaling function $\zeta(q)$ is linear for selfsimilar processes, and  the term {\em multifractal} is normally used only if  the scaling function is strictly concave.
 
Despite the popularity of multifractal models in financial modeling it is difficult to confirm that financial data satisfy multifractal scaling. The structure functions often do not appear to be  well approximated by power laws for high $q$,  mainly because the standard estimators of the structure functions are very unstable for processes with ``fat-tailed'' increment distributions,  unless very long time series are available \cite{Lux:2003vz,Chapman:2005kv}\color{black}. This problem can be partially resolved by supplementing daily data with high-resolution intraday data. For liquid assets the price changes many times per minute, and one can therefore sample the price hundreds of times per day. However, high-frequency financial data contain one-day periodicities which are known to influence the estimated scaling exponents, and therefore these data need to be deseasonalised prior to multifractal analysis. 

It is also possible to improve on the standard estimators by using parametric models and maximal likelihood (ML) methods. For some time, ML estimators have been available for the so-called Markov-Swichting Multifractal (MSM) \cite{Calvet:2004ty}\color{black}, and recently approximate ML methods have also been developed for the Multifractal Random Walk (MRW) \cite{Lovsletten:2011vm,Rypdal:2012ti}\color{black}. The MRW model was constructed by Bacry {\em et al.} \cite{Bacry:2001vc}\color{black}, and it is preferable over the MSM model because of its simplicity. In fact, the multifractality in the MRW model is completely characterised by a single {\em intermittency parameter} $\lambda$. In econometric terms this parameter can also be thought of as a measure of the degree of volatility clustering. The ML estimator for $\lambda$ is very accurate, even for short time series, and can be used as a supplement to the non-parametric estimates of the scaling function. 

In this paper we present an approach which utilises high-frequency data, deseasonalising, and ML estimation in combination with the MRW model. 
As a first example we analyse the share price for a single company in the Oslo Stock Exchange (OSE).
By de-trending intraday data from a one-year period we obtain a time series that is sufficiently long to verify the power-law scaling of the structure functions. The corresponding scaling function is concave and  confirms the  multifractal nature of the stochastic process. We find that the ML estimates of the intermittency parameter in the MRW model are consistent with the results of the structure function analysis, and we can therefore use the ML estimates to quantify the intermittency. The ML method requires less data, and we can obtain meaningful estimates from shorter time records. 

As an application of the methods described above we analyse time-varying intermittency. It has been suggested by several authors to investigate the time-variations of estimated scaling exponents, see e.g. \cite{Carbone:2004jp}\color{black}. Some have even suggested that this type of analysis can be used for prediction of exceptional events, such as crashes and crises  \cite{Los:2006uv}\color{black}. In a recent paper, Morales {\em et al.} \cite{Morales:2012ia} \color{black} calculate structure functions over a weighted moving window to obtain time varying scaling exponents for various share prices, and based on the evolution of these exponents they argue that one can read information about the stability of firms. Although these ideas are intriguing, the statistical significance of the results is yet to be determined. In fact, a simple numerical experiment will demonstrate that, even if applied to synthetic data from a multifractal model with fixed parameters, the moving window approach produces estimates with large time variations. To justify that scaling exponents are changing with time, one must demonstrate that the fluctuations in the exponents estimated from real data are larger than the corresponding variations in synthetic data. Since standard structure functions are so inaccurate, they are not suitable for this type of statistical testing. The situation is improved by applying the methods of this paper. In section \ref{testing} we show that  the intermittency parameter estimated from our example data varies through the year 2008, and peaks in the summer months. The range of this variation is larger than the variability obtained from an ensemble of synthetic realizations of the MRW model, which demonstrates that this result is statistically significant.

The share price analysed here is selected for its high liquidity. We have also chosen a time period where the liquidity is especially high. As a result, the share price changes sufficiently often for us to construct a  time series for which a two-minute resolution is useful. Using ML estimation we compute month-by-month estimates of the intermittency parameter $\lambda$, and we observe that the $\lambda$-curve exhibits a steep fall in the late summer of 2008, at the peak of the financial crisis. This leads to the hypothesis that low values of the intermittency parameter correspond to high market uncertainty. This hypothesis is investigated further by estimating thirteen of the most liquid companies in the OSE  from 2003 to 2009. In this analysis we have used deseasonalised hourly returns, and the estimates are computed for each year of the period. Although there are large variations between the different companies, the average behavior of the intermittency parameter supports our hypothesis. For instance, the year 2008 shows the lowest average value of $\lambda$. We further show that the intermittency parameter is roughly in anti-phase with the investment grade spread in this  period.

Our paper is structured as follows: In section \ref{data} we explain how we we retrieve and deseasonalise the high-frequency data. Standard wavelet-based structure-function analysis is performed. In section \ref{testing} we describe the MRW model and the approximate ML estimator. The ML estimates for our data set is compared with the results of the structure-function analysis, and the statistical significance of the time variation of the intermittency parameter is tested. In section \ref{analysis2} we present a broader analysis for  2003-2009, and make a comparison with the investment grade spread.           
 
\noindent
\begin{figure}[t]
\begin{center}
\includegraphics[width=15.0cm]{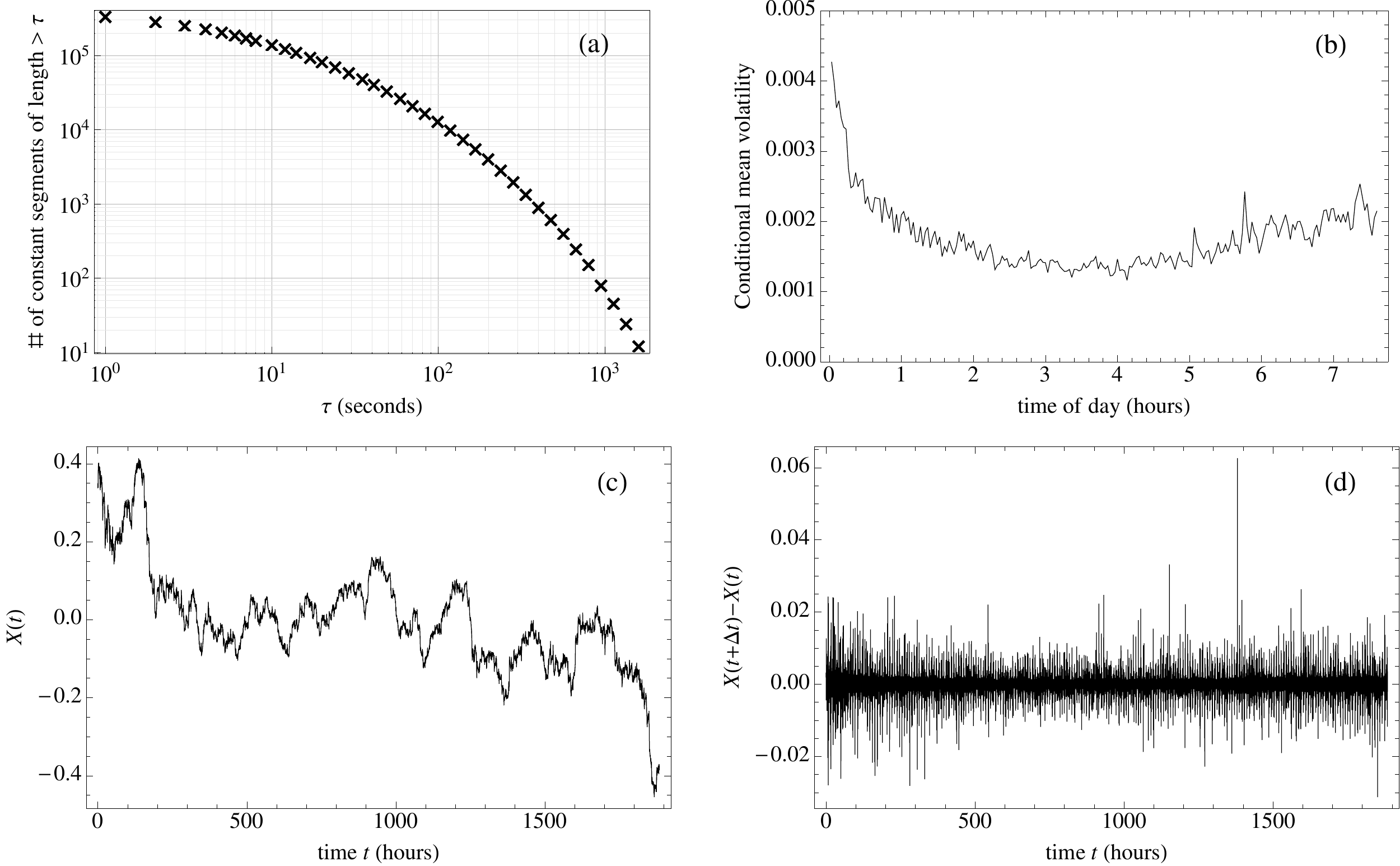}
\caption{(a):  The number of constant-price time intervals longer than $\tau$. (b): Mean absolute value of the log-return conditioned on the time of day. (c): Deseasonalised (logartihmic) price of the REC share. (d): Deseasonalised log-returns, i.e., the increments of the signal in (a). } \label{fig1}
\end{center}
\end{figure}
\noindent
\begin{figure}[t]
\begin{center}
\includegraphics[width=14.0cm]{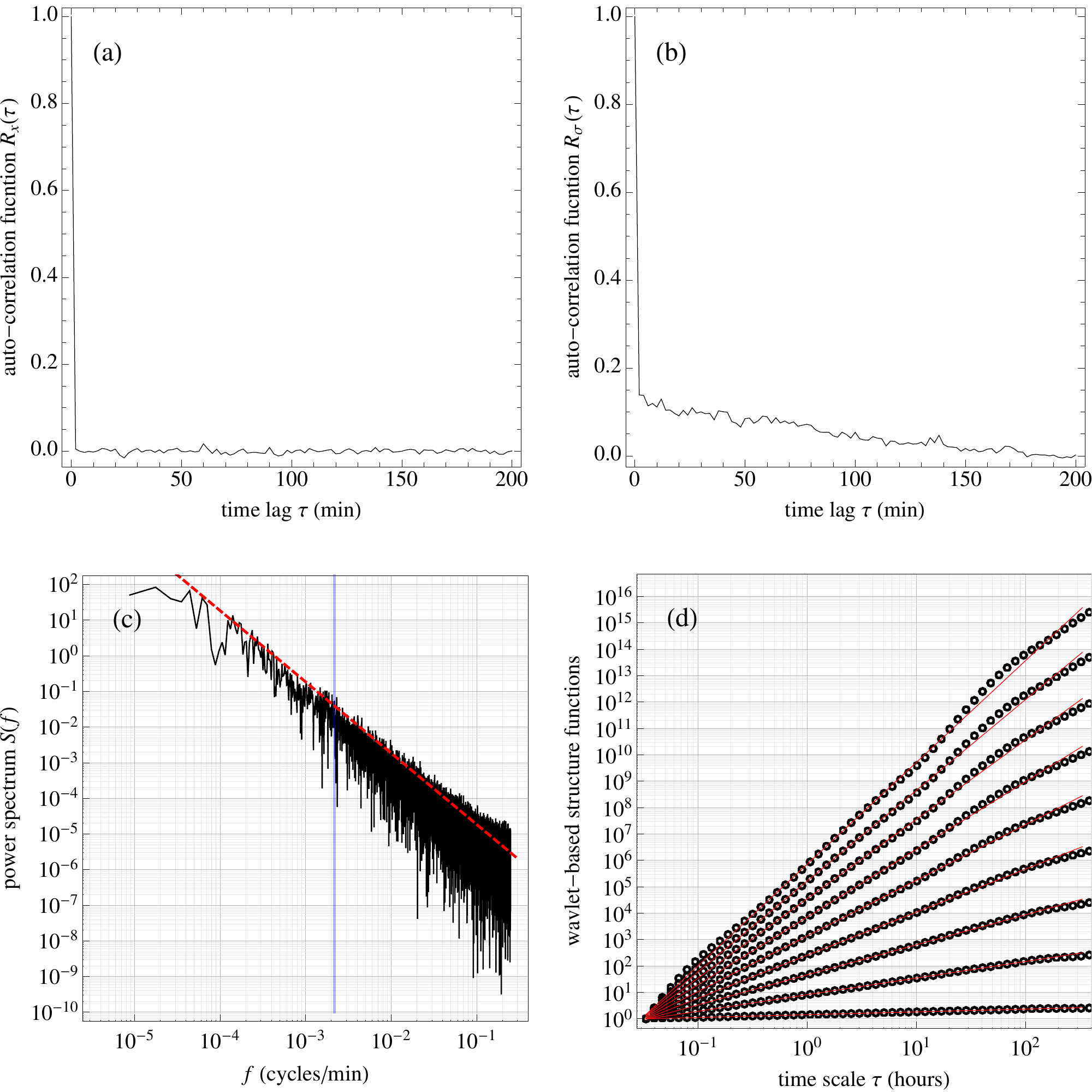}
\caption{(a): The estimated ACF for the deseasonalised log-returns. The dotted lines represent the 95\% percentile for independent data,  $\pm 1.96/\sqrt{n}$, where $n$ is the number of data points in the time series. (b): The estimated ACF for the absolute values of the deseasonalised log-returns. The dotted lines are as in (a). (c): 
The power spectrum of the deseasonalised (logarithmic) price. The dotted line corresponds to a $1/f^2$-spectrum and the vertical line indicates the one-day period. (d): The wavelet-based structure functions of the deseasonalised (logarithmic) price for $q \in \{0.1, 0.6, 1.1,\dots,4.1\}$. The solid lines are the least-square fits that are used to compute the scaling function which is plotted in figure \ref{fig3}(a).  
} \label{fig2}
\end{center}
\end{figure}

\section{Data, deseasonalising and structure-function analysis} \label{data}
The data analysed in this paper is extracted from the \url{stocks.uit.no} database at University of Troms{\o}. The database has been constructed from the OSE data feed in the period 2003-2010, and contains all orders and trades for all instruments available in that period. This makes it possible to reconstruct the complete order book for any security at any point in time. 

For the first example presented in this paper, we have constructed such an order book  for the shares of one company. The time series is sampled for each second of the trading day over one year. If there are no changes in the order book during a second, the order book for the previous second is used. The mid quote of the best bid and ask is then used as a proxy for the market price. For this study liquidity is considered an advantage, so data from the most liquid stock at the exchange, measured in trades per day, is favoured; in this case the stock of Renewable Energy Corporation (REC). We also select the year where this asset has the maximum number of trades,  which is 2008. Hence the data employed in our study is second-by-second mid-quotes for REC for each trading day in 2008. Days when the recorded trading started more than 15 minutes late are removed from the data set (3 trading days), since there is a possibility that such late start might have  beeen caused by errors in the feed. The sample consists of 249 trading days and there are about 5 million changes to the REC order book in the period. That corresponds to about one event every 1.6 seconds on average. However, the mean inter-event time for different trading days varies between 0.5 to 5.6 seconds.

In a preliminary analysis we have divided the time series into constant-price segments and measured the length of these. The distribution of these lengths is shown in figure \ref{fig1}(a). There are about $3 \times 10^5$ such segments, and about one percent  are of length two minutes or more. We therefore sample the data every second minute. This gives 229  data points for each day, and 56 544 data points for the year 2008. We denote this series of log-returns $y_t$. 

The intraday data has a strong one-day periodicity, and this effect is especially strong in the volatility. 
In figure \ref{fig1}(b) we have plotted the mean of $|y_t|$ conditioned on the time-of-day. We observe the so-called ``volatility smile'', with higher volatility in the morning and in the afternoon. We can deseasonalise the periodicity in volatility by dividing each log-return $y_t$ with the mean of $|y_t|$ for the time-of-day corresponding to the time $t$. The resulting time series is denoted $x_t$, and its cumulative sum is denoted $X(t)$, i.e.,
$$
X(t) = \sum_{k=1}^t x_k\,. 
$$      
The signal $X(t)$ is plotted in figure \ref{fig1}(a). 

\begin{rk}
After de-trending the volatility there is still a weak periodicity in the log-returns $x_t$. We deseasonalise this the same way as before; by computing the means of $x_t$ conditioned on the time-of-day and normalising the signal with respect to this curve. The signals  plotted in figures \ref{fig1}(a) and \ref{fig1}(b) have been deseasonalised this way.  
\end{rk}

\begin{figure}[t]
\begin{center}
\includegraphics[width=15.0cm]{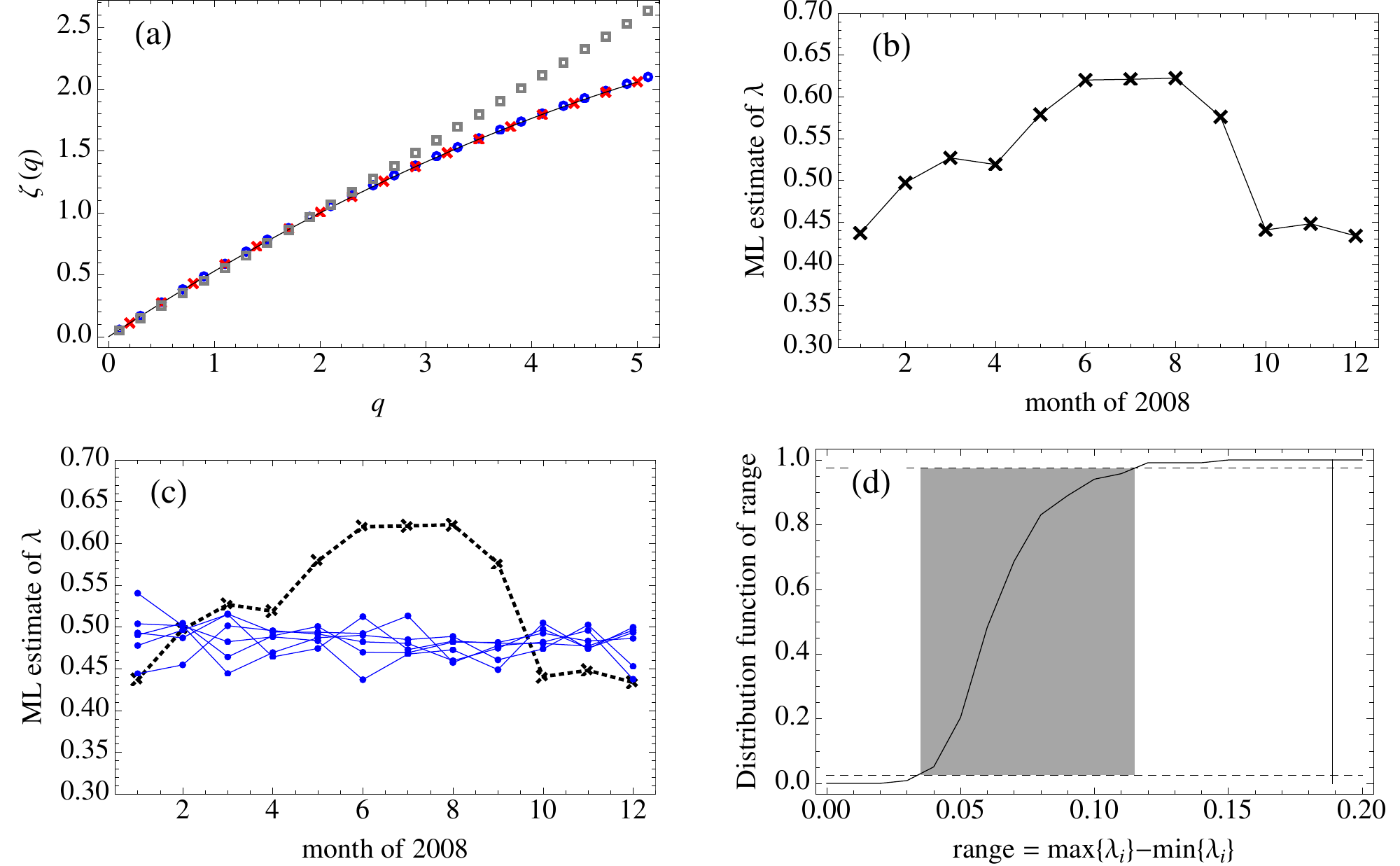}
\caption{(a):  The circles represent the scaling function $\zeta(q)$ obtained from the wavelet-based structure functions which are plotted in figure \ref{fig2}(d). The crosses represent the scaling function estimated from the standard difference-based structure functions. The squares are obtained by applying the wavelet-based structure function analysis to a random walk. The solid curve is the graph of the expression in equation (\ref{zeta}) with $\lambda=0.49$. (b): For each month of 2008 we have plotted the ML estimate of the intermittency parameter $\lambda$ from the REC price data. (c): The dotted curve with crosses is the same as (b), and the full curves with small dots  are computed from realisations of the MRW model with $\lambda=0.5$. (d): For each curve like those plotted  in (c) we compute the range (the difference between the maximum value and the minimum value). The estimated distribution function of this quantity over a large ensemble of realisations is plotted as the solid curve. The shaded region represents the 95\% percentile of this distribution and the solid vertical line represents the range of the curve in (b).} \label{fig3}
\end{center}
\end{figure}

\noindent Figure \ref{fig2}(a) displays the estimated auto-correlation function (ACF) for the signal $x_t$. After deseasonalising, the log-returns appear uncorrelated. In figure \ref{fig2}(b) we have plotted the estimated ACF for the amplitudes $|x_t|$. As opposed to the ACF in figure \ref{fig2}(a), this function decays slowly. 
In figure \ref{fig2}(c) we have estimated the power spectral density (PSD) of the signal $X(t)$. The absence of clear peaks in the PSD indicates that the deseasonalising procedure has been successful. The vertical line represents a one-day period and the dotted line corresponds to a $1/f^2$ spectrum (uncorrelated returns). 
The features described above are typical for multifractal processes, and in figure \ref{fig2}(d) we show the estimated wavelet-based structure functions for the signal $X(t)$. These structure functions are computed by estimating the expectations $\mathbb{E}[|W(t,\tau)|^q]$, where
$$
W(t,\tau) = \frac{1}{\sqrt{\tau}}\int X(t')\, \psi \Big{(} \frac{t-t'}{\tau}\Big{)}\,dt'
$$
is the wavelet transform\footnote{We use a mother wavelet $\psi$ that is the first derivative of a Gaussian.} of $X(t)$. From these moments the scaling function $\zeta(q)$ is defined by the 
relation \cite{Muzy:1991vm}\color{black}: 
\begin{equation} \label{waveletscaling}
\mathbb{E}[|W(t,\tau)|^q] \sim \tau^{\zeta(q)+q/2}.
\end{equation}
Equation (\ref{waveletscaling}) is only meaningful if the wavelet-based structure functions $\mathbb{E}[|W(t,\tau)|^q]$ are well-approximated by power-laws which, as mentioned in the introduction,  is typically difficult to verify in financial time series. However, by employing  deseasonalised high-frequency data, we can demonstrate existence of clear scaling regimes for at least four decades in time and for moments at least as high as $q=4$. This is shown in figure \ref{fig2}(d). The corresponding scaling function is plotted (as circles) in figure \ref{fig3}(a). In this figure we have also included the scaling function estimated from standard difference-based structure function analysis (crosses). The two methods yield very similar results, but he wavelet-based structure functions are better approximated by power-laws, and will therefore be preferred in this study. We have also included the scaling function estimated from a realization of a monofractal random walk (squares) in order to highlight the concave shape of the scaling function for the REC data. The results presented in this section demonstrates the relevance of multifractal methods to these data and serve as verification of the intermittent nature of stock-price fluctuations.

\noindent
\noindent
\begin{figure}[t]
\begin{center}
\includegraphics[width=14.0cm]{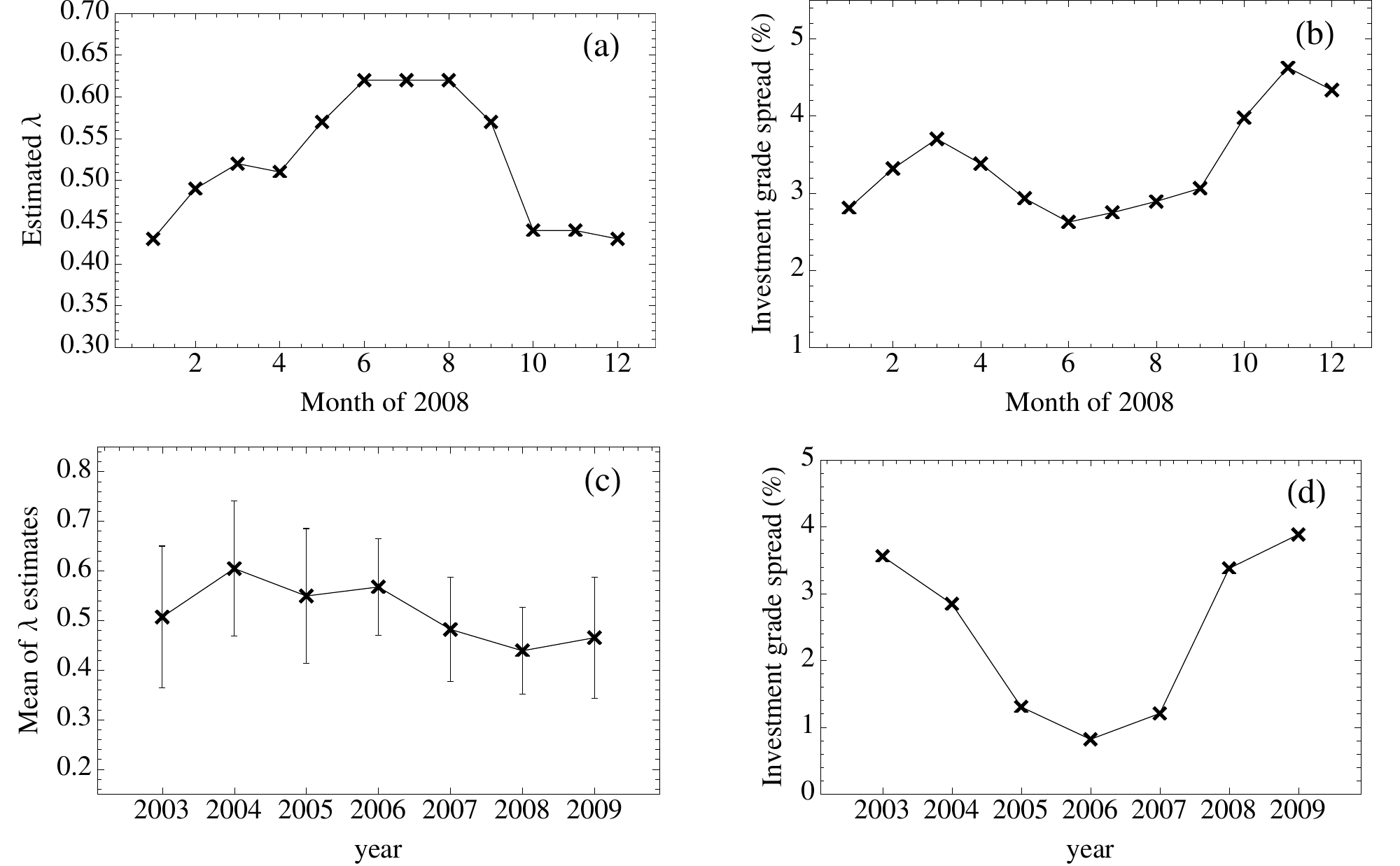}
\caption{(a): Month-by-month ML estimates of the intermittency parameter $\lambda$ in the REC time series (this panel is identical to figure \ref{fig3}(b)). (b): Monthly average of the investment grade spread for the year 2008. (c) The ensemble mean of the year-by-year estimates of $\lambda$ that are presented in table \ref{tab2}. The error bars correspond to the sample standard deviation for each year. (d) Annual averages of the investment grade spread for the years 2003-2009.
} \label{fig4}
\end{center}
\end{figure}

\begin{table}
\begin{center}
\begin{tabular}{||ccccccccccccc||} \hline
Month of 2008 & Jan. & Feb. & Mar. & Apr. & May & Jun. & Jul. & Aug. & Sep. & Oct. & Nov. & Dec. \\ \hline \hline
$\lambda$ & 0.43 & 0.49 & 0.52 & 0.51 & 0.57 & 0.62 & 0.62 & 0.62 & 0.57 & 0.44 & 0.44 & 0.43 \\ \hline 
$\sigma \times 10^3$ & 3.33 & 2.67 & 2.38 & 2.06 & 2.05 & 2.18 & 2.25 & 2.39 & 2.36 & 2.35 & 2.36 & 3.15 \\ \hline \hline
\end{tabular} \caption{ML estimates of $\sigma$ and $\lambda$ computed from the REC data for each month of 2008. The estimates of $\lambda$ are plotted in figure \ref{fig3}(b).} \label{tab1}
\end{center}
\end{table}
\section{Maximum likelihood estimators} \label{testing}
In the previous section, and in figures \ref{fig2} and \ref{fig3}a, we presented non-parametric estimates which confirm the inttermittent nature of stock-price fluctuations. 
In our further analysis we employ a stochastic multifractal model and use an approximate ML method 
to obtain parameter estimates. The model  is a discrete-time approximation to the MRW model of Bacry {\em et al.}  \cite{Bacry:2001vc}\color{black}. Here, the log-returns are expressed as
\begin{equation} \label{MRW}
x_t = \sigma \sqrt{M_t}\,\varepsilon_t\,,
\end{equation}   
where $\varepsilon_t$ is a normalized Gaussian white noise. The stochastic volatility factor is defined by $M_t=c\,e^{h_t}$, where $h_t$ is a centered and stationary Gaussian process with co-variances 
\begin{equation}\label{Cov}
\op{Cov}(h_t,h_s) = \lambda^2 \log^+ \frac{T}{(|t-s|+1) \Delta t}\,.
\end{equation}
The constant $c$ is chosen such that $1/c = \mathbb{E}[e^{h_t}]$. The continuous-time process  approximated by equation (\ref{MRW}) is multifractal, and its scaling function takes a form depending only on the single intermittency parameter $\lambda$ introduced in equation (\ref{Cov});
\begin{equation} \label{zeta}
\zeta(q) = (1+\lambda^2/2)\,q/2-\lambda^2 q^2/8.
\end{equation}
As a preliminary test we have fitted the expression in equation equation (\ref{zeta}) to the wavelet-based estimate of $\zeta(q)$. This fit yields $\lambda = 0.49$, and the corresponding scaling function is shown as the solid curve in figure \ref{fig3}(a). 

A better way to estimate the intermittency parameter is to employ an ML method. 
Given a time series $z$ of length $n$, the ML estimator seeks the parameter vector $\theta=(\lambda,\sigma,T)$ that maximizes the likelihood function $\mathcal{L}_x(\theta|z) = p_x(z|\theta)$. Here $p_x(z|\theta)$ is the $n$-dimensional probability density function (PDF) for the random vector $x=(x_1,\dots,x_n)$, evaluated at the point $z$, given $\theta$. In \cite{Lovsletten:2011vm} \color{black} it is shown how to compute approximations to the likelihood function for the MRW model, and how to use these approximations to estimate parameters. In this work we apply these methods to the high-frequency data of the REC stock price.  

The main advantage of the ML method is that it accurately estimates the intermittency parameter $\lambda$, even for short time series. This allows us to estimate $\lambda$ from a much shorter time record than  required for an estimate from structure function analysis. By employing this method to high-frequency data we obtain meaningful estimates of the intermittency from one month of data. For the deseasonalised REC stock price (with 2 minute time resolution) we have estimated $\lambda$ for each month of the year 2008. The results are presented in table \ref{tab1} and plotted in figure \ref{fig3}(b). The estimated $\lambda$ is higher in the summer months than for the rest of the year which contributes to a range of estimated values this year of $0.19$ for the REC data. 

We need to test the statistical significance of this observed time variability in the estimate of $\lambda$, since it is conceivable that the estimated range of this variation can be attributed to random fluctuations in the estimator.  For such a test we construct an ensemble of 
500 realizations of the MRW model with $\lambda=0.5$. Each realization has the same length as the time series of deseasonalised REC data,  56 544 data points, and each of these realisations is divided in to 12 segments, corresponding to the months of the year. For each realisation we employ the ML method to obtain 12 consecutive estimates of the intermittency parameter. This gives 500 synthetic realizations of the curve in figure \ref{fig3}(b). We have plotted a few of these in figure \ref{fig3}(c). For each of these curves we compute the range (the difference between the maximal and minimal estimate) and construct the distribution of this range. The estimated distribution function is plotted in figure \ref{fig3}(d), together with horizontal  lines corresponding to probabilities 0.025 and 0.975. From the figure one can observe that 95\% of the ranges for the synthetic realisations (the 95\% percentile of the distribution)  are confined to the interval $0.04< \lambda<0.12$. This means that the observed range of 0.19 in the REC data is very improbable given a constant-parameter MRW model.

\begin{table}
\begin{center}
\begin{tabular}{||c||c|c|c|c|c|c|c||} \hline
year & 2003 & 2004 & 2005 & 2006 & 2007 & 2008 & 2009\\ \hline \hline
ACY & 0.78 & 0.64 & 0.49 & 0.54 & 0.530 & 0.46 & 0.37\\
DNB & 0.44& 0.43& 0.42& 0.71 & 0.39 & 0.45 & 0.53 \\
DNO & 0.46& 0.82& 0.56& 0.61& 0.76 & 0.65 & 0.62 \\
FRO & 0.42& 0.60& 0.46& 0.56 & 0.44 & 0.33 & 0.44 \\
NHY & 0.48& 0.40& 0.34& 0.60 & 0.40 & 0.39 & 0.29 \\
NSG & 0.62 & 0.50& 0.89& 0.67 & 0.59 & 0.47 & 0.60 \\
ORK & 0.45& 0.63& 0.64 & 0.43 & 0.39 & 0.37 & 0.32 \\
PSG & 0.70& 0.83& 0.51& 0.46 & 0.42 & 0.36 & 0.36 \\
RCL & 0.46& 0.71& 0.53& 0.61 & 0.55 & 0.40 & 0.45 \\
TAA & 0.35& 0.59& 0.49& 0.61 & 0.48 & 0.50 & 0.64 \\
TEL & 0.26& 0.44& 0.55& 0.39 & 0.42 & 0.39 & 0.45 \\
TGS & 0.59& 0.66 & 0.62& 0.52 & 0.41 & 0.40 & 0.38 \\
TOM & 0.57& 0.59 & 0.64& 0.67 & 0.49 & 0.54 & 0.60 \\ \hline \hline
\end{tabular} \caption{ML estimates of $\lambda$ computed for thirteen of the most liquid firms in the OSE. The estimates are computed for each of the years 2003-2009 based on deseasonalised hourly returns.} \label{tab2}
\end{center}
\end{table}


\section{Comparison with the investment grade spread}  \label{analysis2}
We have shown that the range of variability of estimated $\lambda$ for the REC stock returns cannot be attributed to uncertainty in the estimator. A prominent feature of this variability is the abrupt drop of $\lambda$ in the  months 8-10 of 2008,   at the peak of the 2008 financial crisis (Lehman Brothers filed for bankruptcy on September 15th 2008). This coincidence suggests a connection between market uncertainty and the intermittency parameter of asset prices. In figures \ref{fig4}(a) and \ref{fig4}(b) we compare the the $\lambda$-curve for the REC stock with the investment grade spread. The latter is defined as the difference between the interest rate of AAA bonds and the interest rates on 3-year U.S. federal bonds, and is used as a proxy for the credit spread, and hence as a measure the of perceived market risk. From the figures it appears that the credit spread is roughly in anti-phase with the $\lambda$-curve, at least for the last nine months of 2009, supporting the notion of an anti-correlation between intermittency and market risk.

As a further exploration of this hypothesised connection, it would be useful to compute month-by-month estimates of $\lambda$ for a larger ensemble of stocks, and for a longer period of time. Unfortunately, reliable month-by-month estimates require high liquidity, and are not available for longer time periods for a larger number of stock. We therefore have to resort to year-by-year estimates, making use of time series of deseasonalised hourly returns from thirteen of the most liquid stocks in the OSE. We  present year-by-year estimates of $\lambda$ for each of these for the time period 2003-2009 in table \ref{tab2}. Unfortunately  the estimates vary significantly between the different stocks, precluding  a precise assessment of the evolution of the ensemble averaged $\lambda$ during this period. Nevertheless,   the ensemble mean  yields a $\lambda$-evolution over the period (figure \ref{fig4}(c)) which, taking the large error bars into account, is consistent with an anti-correlation with the investment grade spread shown in figure \ref{fig4}(d). In particular, we observe that the investment grade spread increases from the middle towards the end of the decade, while  the ensemble-averaged $\lambda$ decreases during this time.  

\section{Conclusion}
The main result of this paper is that the combination of high-frequency data and deseasonalising provides sufficient information to estimate the intermittency parameter in stock prices and to confirm the multifractal nature of these   signals. We also show that if we supplement the structure-function analysis with the MRW model and its corresponding ML estimator, we can estimate the intermittency from short time records (a month). For the data considered in this paper it is observed that the estimated intermittency varies in time, and we have concluded that this variation is statistically significant. One implication is that it is not possible to describe the observed data using the MRW model with constant parameters, and it therefore lends meaning to the notion of time-varying degree of intermittency. 

Based on the examples presented in this paper, we suggest  that there is a negative correlation between uncertainty in the financial market and the intermittency of stock price fluctuations. This hypothesis will be investigated further in future work, where the techniques presented in this paper will be applied in a broader study of time-variations of estimated scaling exponents.


\end{document}